%% file: he_transition_v9.tex
\documentclass[twocolumn,aps,prl,floatfix]{revtex4-1}
\usepackage{graphicx}
\usepackage{times}
\usepackage{nicefrac}
\usepackage{amsmath}
\usepackage{amsfonts}
\usepackage{amssymb}
\usepackage{amsthm}
\usepackage{epsf}
\usepackage{bm}
\usepackage{times}

\usepackage{dcolumn}
\usepackage{siunitx}

\newcommand{\be}{\begin{eqnarray}}
\newcommand{\ee}{\end{eqnarray}}

\usepackage{ulem}
\usepackage{color}
\definecolor{BLUE}{rgb}{0.0,0.0,1.0}

\begin{document}

\title{QED calculations of the $n=2$ to $n=1$ x-ray transition energies in middle-$Z$ heliumlike ions}

\author{A. V. Malyshev}
\affiliation {Department of Physics, St.~Petersburg State University, 
Universitetskaya 7/9, 199034 St.~Petersburg, Russia}

\author{Y. S. Kozhedub}
\affiliation {Department of Physics, St.~Petersburg State University, 
Universitetskaya 7/9, 199034 St.~Petersburg, Russia}

\author{D. A. Glazov}
\affiliation {Department of Physics, St.~Petersburg State University, 
Universitetskaya 7/9, 199034 St.~Petersburg, Russia}

\author{I. I. Tupitsyn}
\affiliation {Department of Physics, St.~Petersburg State University, 
Universitetskaya 7/9, 199034 St.~Petersburg, Russia}

\author{V. M. Shabaev}
\affiliation {Department of Physics, St.~Petersburg State University, 
Universitetskaya 7/9, 199034 St.~Petersburg, Russia}

\begin{abstract}

  \textit{Ab initio} QED calculations of the four x-ray transitions from the $L$ to $K$ shell in heliumlike argon, titanium, iron, copper, and krypton are performed. The binding energies for all the $n=1$ and $n=2$ states are evaluated as well. The calculation approach combines the rigorous QED treatment in the first two orders of the perturbation theory constructed within the extended Furry picture with the third- and higher-order correlation effects evaluated in the Breit approximation. The obtained results are compared with the previous evaluations and available experimental data.

\end{abstract}

\maketitle

\section{Introduction}

Bound-state quantum electrodynamics (QED) has proved to be an efficient tool to predict the binding and transition energies in highly charged ions, see Refs.~\cite{ Schweppe:1991:1434, Stoehlker:2000:3109, Brandau:2003:073202, Draganic:2003:183001, Gumberidze:2004:203004, Gumberidze:2005:223001, Beiersdorfer:2005:233003, Mackel:2011:143002} for the corresponding high-precision measurements and Refs.~\cite{ Persson:1996:204, Yerokhin:1997:361, Yerokhin:2006:253004, Artemyev:2007:173004, Kozhedub:2008:032501, Kozhedub:2010:042513, Sapirstein:2011:012504, Artemyev:2013:032518, Yerokhin:2015:033103, Yerokhin:2018:052509} for the related theory. Heliumlike ions represent the simplest few-electron systems where many-electron relativistic and QED effects can be studied. 
There are many systematic theoretical investigations of He-like ions \cite{Drake:1988:586, Johnson:1992:R2197, Chen:1993:3692, Plante:1994:3519, Cheng:1994:247, Indelicato:1995:1132, Cheng:2000:044503}.
To date, the most elaborated treatment of the energies of ground and low-lying excited states in heliumlike highly charged ions has been performed in Ref.~\cite{Artemyev:2005:062104}, where the second-order two-electron QED contributions were rigorously evaluated to all orders in $\alpha Z$ ($\alpha$ is the fine structure constant and $Z$ is the nuclear charge number) for $Z\geqslant 12$.
The theoretical predictions from Ref.~\cite{Artemyev:2005:062104} were generally in good agreement with the available at that time experimental data for the $K\alpha$~\cite{Deslattes:1984:L689,Beiersdorfer:1989:150,Chantler:2000:042501} as well as intra-$L$-shell~\cite{Howie:1994:4390,Kukla:1995:1905,Redshaw:2001:23002} transition energies, for more detailed comparison see Table~VIII in Ref.~\cite{Artemyev:2005:062104}.

A new great interest to 
middle-$Z$ heliumlike ions has been triggered
 by Chantler and co-authors \cite{Chantler:2012:153001,Chantler:2014:123037}. In Ref.~\cite{Chantler:2012:153001}, based on the high-precision measurement of x-ray resonance line in He-like titanium and a statistical analysis of the previous experimental data, it was claimed that there is a discrepancy between experiment and the theoretical predictions obtained by Artemyev {\textit{et al.}}~\cite{Artemyev:2005:062104} which grows approximately as~$Z^3$. As a response, a number of new measurements of the x-ray transitions in heliumlike ions has followed~\cite{Rudolph:2013:103002, Schlesser:2013:022503, Kubicek:2014:032508, Epp:2015:020502_R, Beiersdorfer:2015:032514, Machado:2018:032517}. New statistical analyses based on the extended sets of experimental data have been undertaken~\cite{Epp:2013:159301, Beiersdorfer:2015:032514, Machado:2018:032517}. These investigations have refuted the aforementioned $Z^3$-deviation trend. Nevertheless, in view of the general scatter of the experimental values the issue of independent \textit{ab initio} calculations of the transition energies in heliumlike ions  has become very urgent. 

There are some points in Ref.~\cite{Artemyev:2005:062104} that can be improved employing the most advanced methods which are available to date.
In our recent work~\cite{Malyshev:2018:085001} we revisited the contribution of the nuclear recoil effect to the low-lying energy levels of two-electron ions. Having treated this effect in a more accurate way and compared with the results from Ref.~\cite{Artemyev:2005:062104}, we have shown that the nuclear recoil contribution can not be responsible for the discrepancy declared by Chantler \textit{et al.}. 
In the present work we 
address all the other contributions and
perform from scratch the rigorous QED calculations of the $n=2$ to $n=1$ transition energies in middle-$Z$ heliumlike ions within the extended Furry picture. Namely, the x-ray transitions from levels $1s2p \,^1P_1$, $1s2p \,^3P_2$, $1s2p \,^3P_1$, and $1s2s \,^3S_1$ to the $1s^2 \, {}^1S_0$ ground state, which are traditionally denoted as $w$, $x$, $y$, and $z$~\cite{Gabriel:1972:99}, respectively, are evaluated for heliumlike argon ($Z=18$), titanium ($Z=22$), iron ($Z=26$), copper ($Z=29$), and krypton ($Z=36$).
The obtained results are compared with the theoretical predictions from Ref.~\cite{Artemyev:2005:062104} and available experimental values.


\section{Theoretical methods}

The Furry picture of QED is generally recognized as a good starting point for description of highly charged ions. To zeroth order, within this picture the interelectronic interaction is neglected and the electronic states in the presence of the nuclear Coulomb potential $V_{\rm nucl}$ are considered. This means that within the initial approximation electrons obey the one-electron Dirac equation. The interaction with other electrons and with the quantized electromagnetic field are treated by the QED perturbation theory which can be formulated employing, e.g., the two-time Green's function (TTGF) method~\cite{TTGF}. It is precisely this approach that was used in Ref.~\cite{Artemyev:2005:062104} to evaluate the ionization energies of He-like ions. 

Nowadays, the state-of-the-art methods of bound-state QED are limited by the consideration of the second-order contributions. 
Therefore, it is impossible to account for the higher-order interelectronic-interaction corrections to the Lamb shift rigorously in the framework of bound-state QED. 
However, one can do it partially by employment of the extended version of the Furry picture.
In the extended Furry picture the nuclear potential is replaced with some effective potential: $V_{\rm nucl} \rightarrow V_{\rm eff} = V_{\rm nucl} + V_{\rm scr}$, where the local screening potential $V_{\rm scr}$ models the interaction of electrons. Obviously, one has to account for the counterterm $\delta V = -V_{\rm scr}$, that leads to new Feynman diagrams to be evaluated.
The extended version of the Furry picture was successfully applied to the high-precision QED calculations of the atomic properties in highly charged ions \cite{Sapirstein:2001:022502, Artemyev:2007:173004, Yerokhin:2007:062501, Kozhedub:2010:042513, Sapirstein:2011:012504, Artemyev:2013:032518, Malyshev:2017:022512, Sapirstein:2001:032506, Glazov:2006:330, Volotka:2014:253004} and many-electron atoms~\cite{Sapirstein:2002:042501, Chen:2006:042510, Sapirstein:2003:022512, Sapirstein:2006:042513}. 

An additional difficulty from the theoretical point of view is that the $1s2p\, ^1P_1$ and $1s2p\, ^3P_1$ states are quasidegenerate. The treatment of such states is considerably more complicated than in case of a single level. To describe this pair of quasidegenerate states, the TTGF method involves the evaluation of a $2\times 2$ matrix $H$ which should include all the relevant contributions~\cite{TTGF}. The matrix $H$ is calculated by the perturbation theory starting from the unperturbed states $(1s2p_{1/2})_1$ and $(1s2p_{3/2})_1$ defined in the $jj$-coupling. The total binding energies of the $1s2p\, ^1P_1$ and $1s2p\, ^3P_1$ states can be determined by diagonalizing the matrix $H$. This scheme was realized first in Refs.~\cite{Artemyev:2005:062104,Artemyev:2000:022116}. Alternative approaches to some of the second-order contributions were considered in Refs.~\cite{Lindgren:2001:062505,Andreev:2004:062505}.

\begin{figure}
\begin{center}
\includegraphics[width=0.89\linewidth]{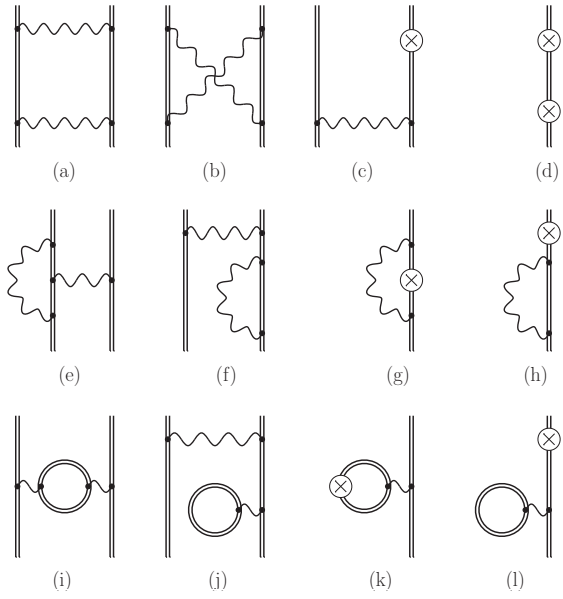}
\caption{\label{fig:feyn_2el}
Second-order two-electron Feynman diagrams and corresponding counterterm diagrams arising in the extended Furry picture. 
A double line denotes the electron propagator in the binding potential. A wavy line corresponds to the photon propagator.
A circle with a cross represents the counterterm $\delta V$. 
}
\end{center}
\end{figure}

In the present work, the formalism applied in Ref.~\cite{Artemyev:2005:062104} for \textit{ab initio} calculations of the second-order two-electron QED contributions (Fig.~\ref{fig:feyn_2el}, the first and second columns) has been thoroughly revised and extended to the case when a screening potential is incorporated into the zeroth-order approximation.
The corresponding one- and two-electron counterterm diagrams are shown in the last two columns of Fig.~\ref{fig:feyn_2el}.
We have performed the calculations with the use of two different types of effective potential. The core-Hartree (CH) and local Dirac-Fock (LDF) screening potentials were employed in order to partially account for the higher-order electron-electron interaction effects. The construction
procedures and applications for these potentials can be found, e.g.,
in Refs.~\cite{Sapirstein:2002:042501, Shabaev:2005:062105, Malyshev:2017:022512}. Moreover, in order to examine the accuracy
of the results obtained in Ref.~\cite{Artemyev:2005:062104}, we have performed the calculations within the standard Furry
picture for the Coulomb potential as well. 


\begin{figure}
\begin{center}

\begin{minipage}[c]{0.001\linewidth}
\scriptsize
(a)
\vspace{5mm}
\end{minipage}
\hfill
\begin{minipage}{0.985\linewidth}
\includegraphics*[width=0.978\linewidth]{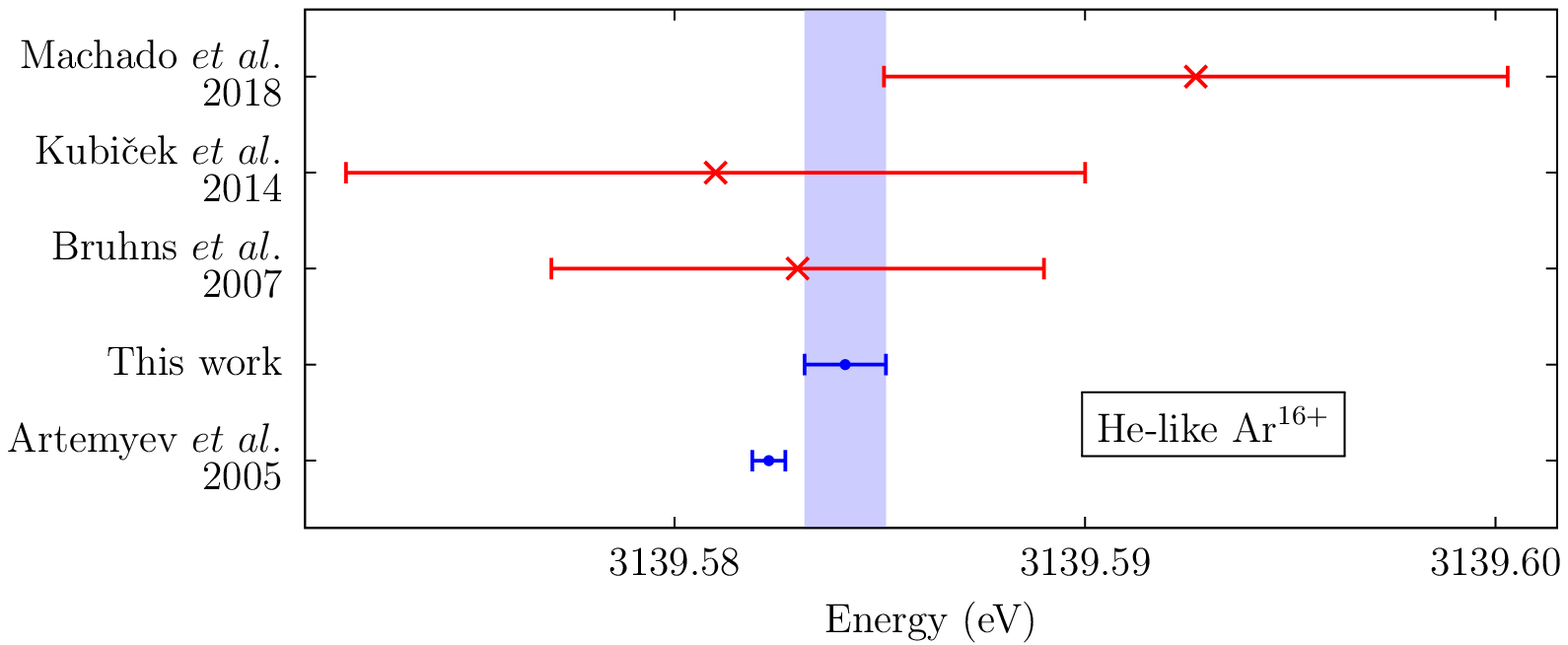}
\end{minipage}

\vspace*{1mm}

\begin{minipage}[c]{0.001\linewidth}
\scriptsize
(b)
\vspace{5mm}
\end{minipage}
\hfill
\begin{minipage}{0.985\linewidth}
\includegraphics*[width=0.978\linewidth]{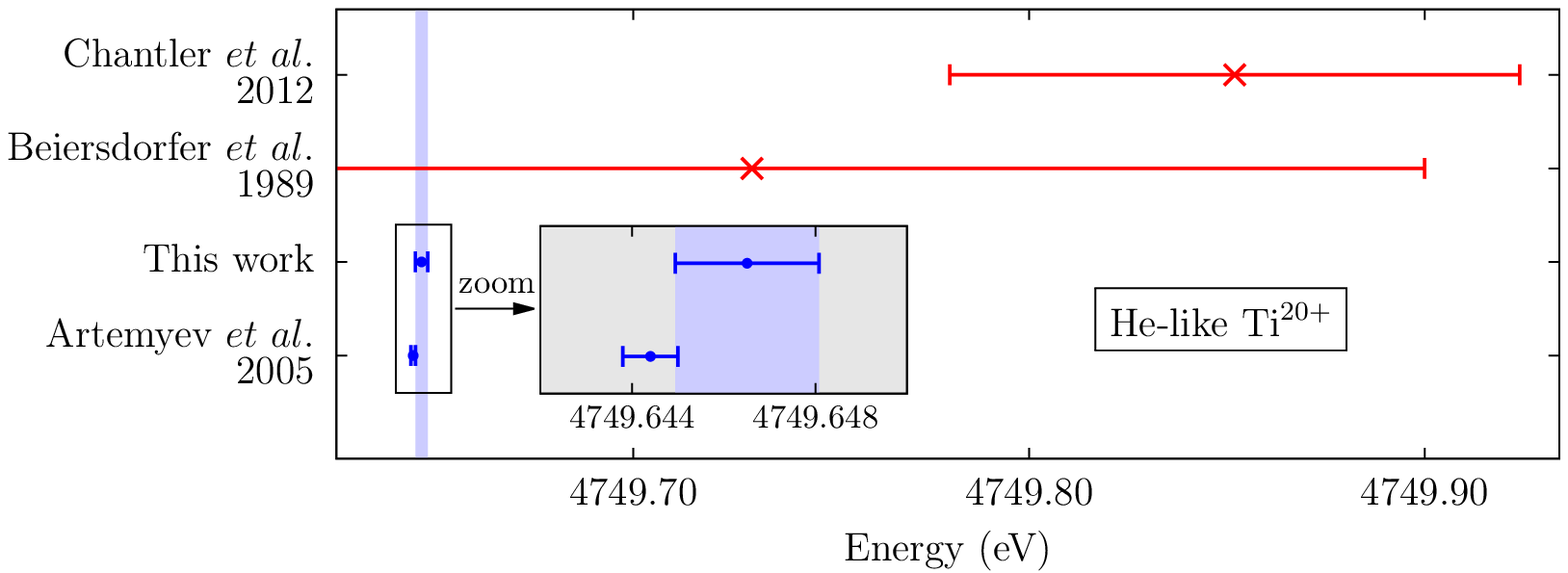}
\end{minipage}

\vspace*{1mm}

\begin{minipage}[c]{0.001\linewidth}
\scriptsize
(c)
\vspace{5mm}
\end{minipage}
\hfill
\begin{minipage}{0.985\linewidth}
\includegraphics*[width=0.978\linewidth]{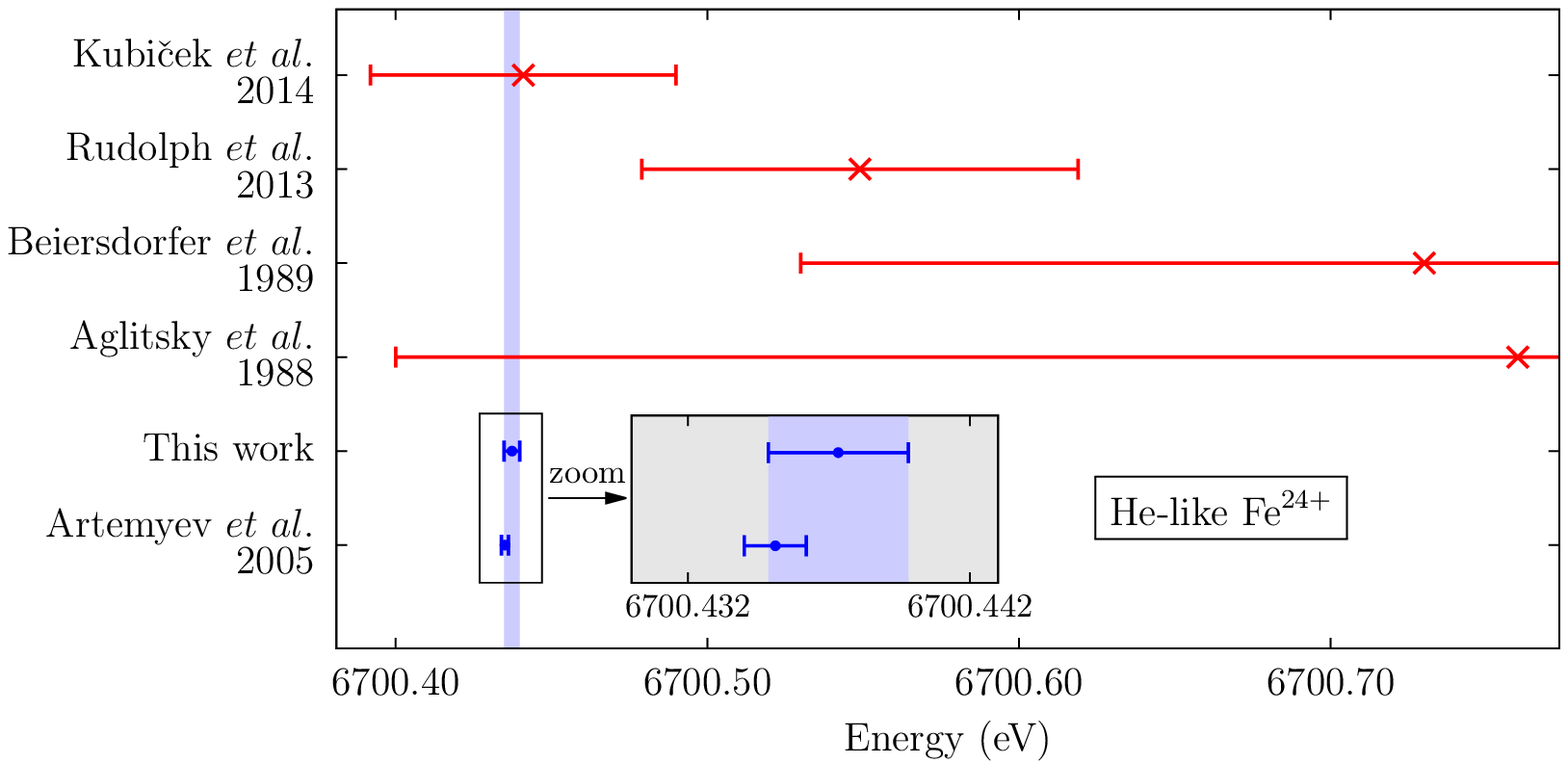}
\end{minipage}

\vspace*{1mm}

\begin{minipage}[c]{0.001\linewidth}
\scriptsize
(d)
\vspace{5mm}
\end{minipage}
\hfill
\begin{minipage}{0.985\linewidth}
\includegraphics*[width=0.978\linewidth]{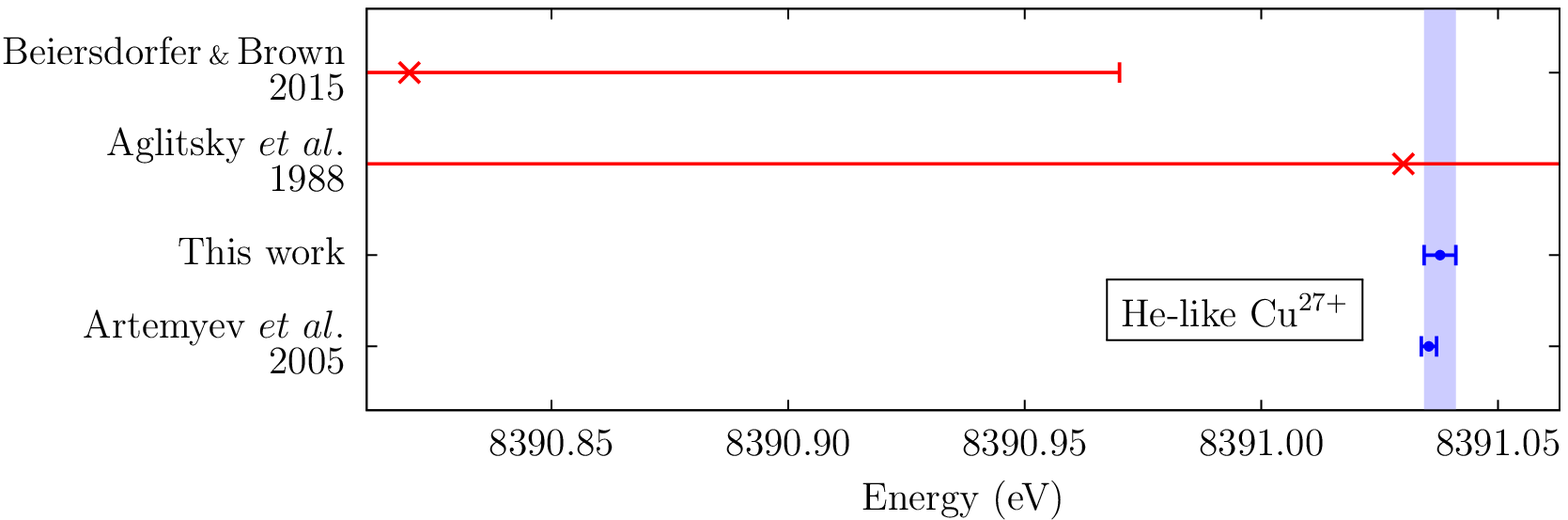}
\end{minipage}

\vspace*{1mm}

\begin{minipage}[c]{0.001\linewidth}
\scriptsize
(e)
\vspace{5mm}
\end{minipage}
\hfill
\begin{minipage}{0.985\linewidth}
\includegraphics*[width=0.978\linewidth]{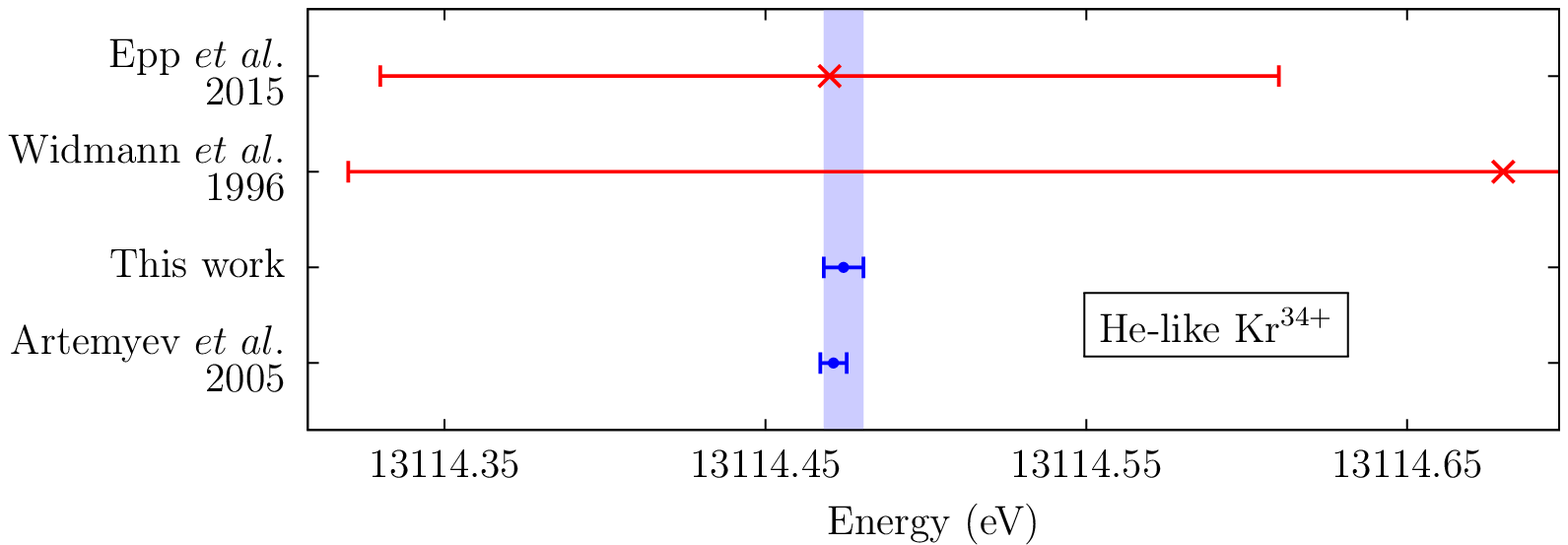}
\end{minipage}

\caption{\label{fig:w_line}
The $w(1s2p\,^1P_1 \rightarrow 1s^2\,^1S_0)$-line transition energy in He-like ions: (a) argon; (b) titanium; (c) iron; (d) copper; (e) krypton. Our theoretical prediction and the result by Artemyev \textit{et al.}~\cite{Artemyev:2005:062104}, reevaluated using the CODATA 2014 recommended values of the fundamental constants~\cite{Mohr:2016:035009}, are given as blue dots. The theoretical uncertainty of the present calculations is shown as a blue area. The selected experimental values are given as red crosses and are from the following sources: ${\rm Ar}^{16+}$, Refs.~\cite{Bruhns:2007:113001, Kubicek:2014:032508, Machado:2018:032517}; ${\rm Ti}^{20+}$, Refs.~\cite{Beiersdorfer:1989:150, Chantler:2012:153001}; ${\rm Fe}^{24+}$, Refs.~\cite{Aglitsky:1988:136, Beiersdorfer:1989:150, Rudolph:2013:103002, Kubicek:2014:032508}; ${\rm Cu}^{27+}$, Refs.~\cite{Aglitsky:1988:136, Beiersdorfer:2015:032514}; ${\rm Kr}^{34+}$, Refs.~\cite{Widmann:1996:2200, Epp:2015:020502_R}.
}
\end{center}
\end{figure}


Consideration of the two-loop one-electron diagrams completes the QED treatment of the energy levels in He-like ions
to second order in $\alpha$ within the Furry picture~\cite{Yerokhin:2003:203, Yerokhin:2006:253004, Yerokhin:2018:052509}. 
\textit{Ab initio} calculation of these contributions to all orders in $\alpha Z$ is a very challenging problem.
The breakthrough in this field is represented by the all-order calculations of the two-loop self-energy corrections by Yerokhin \textit{et al.}~\cite{Yerokhin:2003:203, Yerokhin:2006:253004, Yerokhin:2018:052509}.
However, some two-loop diagrams are still evaluated only within the free-loop approximation~\cite{Yerokhin:2008:062510}.
To account for these contributions
we use the results compiled in Ref.~\cite{Yerokhin:2015:033103}.

The evaluation in the Coulomb potential of the contributions corresponding to the diagrams in Figs.~\ref{fig:feyn_2el}(a) and (b) allows one to treat the correlation effects to second order in $1/Z$ rigorously within QED. Extending the calculations to the case with a screening potential included into the initial approximation [this implies consideration of the counter\-term diagrams in Figs.~\ref{fig:feyn_2el}(c) and (d)] provides an opportunity to cover the higher-order corrections partly. Nevertheless, the remaining part of the interelectronic interaction has to be taken into account at least within the lowest-order relativistic approximation. In Ref.~\cite{Artemyev:2005:062104} the issue to evaluate the correlations effects of third and higher orders in $1/Z$ was addressed by summing terms of the $1/Z$ expansion. The corresponding coefficients were taken from Refs.~\cite{Sanders:1969:84, Aashamar:1970:3324} for nonrelativistic energies and from Ref.~\cite{Drake:1988:586} for the Breit-Pauli correction. Within the extended Furry picture this approach is no longer valid. 
Therefore, for all the potentials we evaluate this contribution on the basis of the Dirac-Coulomb-Breit Hamiltonian employing the two independent methods. 
The first one is the relativistic configuration-interaction (CI) method based on Dirac-Sturm orbitals~\cite{Tupitsyn:2003:022511}. The procedure how to extract the desired higher-order interelectronic-interaction contribution from the total CI result for the case of a single level was discussed, e.g., in Refs.~\cite{Kozhedub:2010:042513, Artemyev:2013:032518, Malyshev:2017:022512}.
In the present work, this procedure has been generalized to deal with the quasidegenerate states.
The second method directly yields the third- and higher-order contributions by means of the recursive perturbation theory. The description of this approach for a single-level case can be found in Refs.~\cite{Glazov:2017:46, Malyshev:2017:022512}. In order to generalize the perturbation theory to the case of the quasidegenerate states, we used the combinatorial algorithm proposed in Ref.~\cite{Brouder:2012:2256}. The results obtained employing these two different approaches are found to be in good agreement. The details of both methods will be published elsewhere.

Finally, one has to account for the nuclear recoil effect which lies beyond the external-field approximation, that is beyond the Furry picture. The detailed analysis of the nuclear recoil contribution to the energy levels in He-like ions was presented in Ref.~\cite{Malyshev:2018:085001}. We have extended this scheme to include the screening potential into the unperturbed Hamiltonian. 
Moreover, to be consistent with the most recent compilation of the Lamb shift in H-like ions~\cite{Yerokhin:2015:033103}, we have taken into account the small correction due to the reduced-mass dependence in the one-electron self-energy, vacuum-polarization, and two-loop QED contributions, see Ref.~\cite{Yerokhin:2015:033103} for details.


\input{tables_v4.tex}


All the calculations have been performed employing the Fermi model for the nuclear charge distribution. The most abundant isotopes are selected, and the nuclear radii are taken from Ref.~\cite{Angeli:2013:69}. 

\section{Numerical results and discussion}

As noted above, in our calculations the zeroth-order approximation is given by the two-electron wave functions
constructed in the $jj$-coupling from the solutions of the one-electron Dirac equation. We study all the states $(1s\,nlj)_J$,
where $nlj=1s,2s,2p_{1/2},2p_{3/2}$ and $J$ is the total angular momentum.
In order to keep better under control the accuracy of the calculations, we have performed them
with the Coulomb, LDF, and CH potentials.
The individual contributions to the binding energies of He-like argon
are presented in Table~\ref{tab:Ar_contrib}.
For the mixing configurations, $(1s2p_{1/2})_1$ and $(1s2p_{3/2})_1$ denote the diagonal $H$-matrix elements, while ``off-diag.'' stands for the off-diagonal contributions.
The  $\Delta E_{\rm Dirac}$ value is the energy obtained from the Dirac equation.
The first- and second-order interelectronic-interaction contributions calculated employing the rigorous QED approach and the higher-order correlation corrections evaluated within the Breit approximation 
are given by  $\Delta E_{\rm int}^{(1)}$, $\Delta E_{\rm int}^{(2)}$, and $\Delta E_{\rm int}^{(\ge 3)}$,
respectively. The terms $\Delta E_{\rm 1el}^{\rm QED}$ and  $\Delta E_{\rm 1el}^{\rm ScrQED}$ denote
the one-electron and screened QED contributions, respectively. Finally,  $\Delta E_{\rm rec}$ is
the recoil correction. 
The total values obtained for all the potentials are presented in the last column. For the calculations with the Coulomb potential two total values are given. The second one includes the higher-order QED correction $\Delta E_{\rm ho}^{\rm QED}$ evaluated using the related formula from Ref.~\cite{Drake:1988:586}. Since the latter correction was obtained for the Coulomb potential it can be included only within the standard Furry picture. It is seen that while the individual terms in Table~\ref{tab:Ar_contrib} may differ from line to line, the total values of the binding energies obtained for the different potentials are in good agreement.

In the case of the Coulomb potential,
 the individual contributions from Table~\ref{tab:Ar_contrib} have been compared with
the corresponding values obtained by Artemyev \textit{et al.}~\cite{Artemyev:2005:062104}.
It was found that there exists
a small discrepancy between the results for the two-photon exchange
and higher-order correlation contributions.
This discrepancy is most pronounced for the ground $1s^2$ state, resulting in a slight systematic shift of our theoretical predictions for the $L$ to $K$ transition energies compared to the results of Ref.~\cite{Artemyev:2005:062104}, see the discussion below.
We note also that
some additional small discrepancy 
was caused by the employment of
the different values of the fundamental constants (in the present work the CODATA 2014 recommended values~\cite{Mohr:2016:035009} were used: $\alpha^{-1}=137.035\,999\,139(31)$ and $m_{\rm e}c^2=0.510\,998\,9461(31)$~MeV).

From Table~\ref{tab:Ar_contrib}, one can see that the addition of the correction $\Delta E_{\rm ho}^{\rm QED}$, evaluated according to Ref.~\cite{Drake:1988:586}, to the Coulomb results generally brings them closer to the calculations performed using the effective potentials. However, it is not always the case. For instance, we found that  with increasing the nuclear charge number $Z$ the higher-order correction $\Delta E_{\rm ho}^{\rm QED}$ for the ground $1s^2$ state tends to shift the Coulomb value in the opposite direction enlarging the difference between the results for the Coulomb and effective potentials. The same trend was noticed in Ref.~\cite{Holmberg:2015:012509}. 
To illustrate this for the ground-state binding energy of heliumlike ions, in the second and third columns of Table~\ref{tab:ho} we show the difference between the results obtained for the effective potentials and the Coulomb potential. In the fourth column the correction $\Delta E_{\rm ho}^{\rm QED}$ is given, whereas in the last column the combined correlation and QED shift evaluated in Ref.~\cite{Holmberg:2015:012509} is presented. Although not exactly equivalent, these three approaches serve to estimate the higher-order QED contribution. One can see that our results are generally in agreement with the data presented in Ref.~\cite{Holmberg:2015:012509} and increase monotonically with $Z$, while the higher-order correction $\Delta E_{\rm ho}^{\rm QED}$ calculated according to Ref.~\cite{Drake:1988:586} changes the sign as $Z$ grows.


The results for the LDF potential 
have been used as the final values. 
Our theoretical predictions for the binding energies 
for all the $n=1$ and $n=2$ states in heliumlike argon, titanium, iron, copper, and krypton 
are shown in Table~\ref{tab:binding}. The energies of the $1s2p\,^1P_1$ and $1s2p\, ^3P_1$ states are obtained by diagonalizing the matrix~$H$ with the elements given by the total values from the rows $(1s2p_{1/2})_1$, $(1s2p_{3/2})_1$, and ``off-diag.'' in Table~\ref{tab:Ar_contrib}. The uncertainties were estimated by summing quadratically several contributions. The first one in addition to the numerical uncertainties of the second-order two-electron terms includes the uncertainty due to the nuclear size effect and the uncertainty due to uncalculated two-loop one-electron QED corrections~\cite{Yerokhin:2015:033103}. The second uncertainty is associated with the uncalculated higher-order screening QED contributions. 
Firstly, it was estimated considering the scatter of the total values obtained in the calculations with the different potentials. 
If one took into account the correlation and QED corrections to all orders, the total results would be fully independent of the choice of initial approximation. 
Therefore, the discrepancy between the different calculations provides the estimation of the uncalculated terms. 
For the excited states we took the maximum between the scatter for this state and the scatter for the ground state divided by the factor of four which conservatively describes the scaling of the QED effects for these states. Secondly, taking into account that the interelectronic interaction for highly charged ions can be treated within the $1/Z$ perturbation theory we estimated the uncalculated higher-order QED corrections by multiplying the two-loop one-electron QED term for the $1s$ state by the conservative factor $2/Z$.


The $n=2$ to $n=1$ x-ray transition energies for the $w$, $x$, $y$, and $z$ lines are given in Table~\ref{tab:wxyz}. They were obtained by subtracting the ground-state binding energy from the binding energies of the excited states. 
In Table~\ref{tab:wxyz} our theoretical predictions are compared with the results of the previous calculations. Compared to the calculations by Artemyev \textit{et al.}~\cite{Artemyev:2005:062104}, our estimation of the uncertainty due to the uncalculated higher-order QED corrections has been performed in a more conservative way. There is a slight discrepancy with the results from Ref.~\cite{Artemyev:2005:062104} for heliumlike argon and titanium. We suppose that it is caused by an underestimation of the uncertainty in the calculations performed in Ref.~\cite{Artemyev:2005:062104}. In addition, the comparison with available experimental values is given in Table~\ref{tab:wxyz}. Our results are in agreement with the most recent measurements~\cite{Machado:2018:032517, Kubicek:2014:032508, Rudolph:2013:103002, Beiersdorfer:2015:032514, Epp:2015:020502_R} of the x-ray transitions in middle-$Z$ heliumlike ions. The $w(1s2p\,^1P_1 \rightarrow 1s^2\,^1S_0)$-line transition energy is plotted with selected experimental values in Figs.~\ref{fig:w_line}(a)-(e). 
For convenience, the smaller energy regions, encompassing both theoretical values, are zoomed for heliumlike titanium and iron. 
As one can see from Table~\ref{tab:wxyz} and Fig.~\ref{fig:w_line}(b), even though \textit{ab initio} QED calculations have been performed with the most advanced methods and all the possible sources of the theoretical uncertainty have been studied, the discrepancy between theory and the measurement by Chantler \textit{et al.}~\cite{Chantler:2012:153001} for He-like titanium still persists. 

Finally, since it was being discussed in the literature~\cite{Epp:2015:020502_R}, for heliumlike krypton we present our theoretical prediction for the ratio of the $w$-line transition energy to the $y$-line transition energy, $[E(w)/E(y)]_{\rm Th}=1.006\,7828(7)$. This value is matching exactly with the one which can be derived from the results by Artemyev \textit{et al.}~\cite{Artemyev:2005:062104} and in good agreement with the value measured in Ref.~\cite{Epp:2015:020502_R}, $[E(w)/E(y)]_{\rm Exp}=1.006\,780(7)$. This ratio gives the value of the $w$-line in energy units of the $y$-line independent of any energy calibration and can be measured to a high precision. To date, the theoretical precision for this ratio in krypton is one order of magnitude higher than the experimental one. The accuracy of the theoretical prediction can be further improved provided the more precise measurement will be performed.

\section{Conclusions}

To summarize, we have performed \textit{ab initio} QED calculations of the binding energies for all the $n=1$ and $n=2$ states in middle-$Z$ He-like ions with the most advanced methods available to date.
The $n=2$ to $n=1$ x-ray transition energies are determined. The calculations merge the rigorous QED treatment in the first and second orders of the perturbation theory and the higher-order interelectronic-interaction effects evaluated within the Breit approximation.
The obtained x-ray transition energies are generally in good agreement with the most recent high-precision measurements for heliumlike ions. From the theoretical side, no possible explanations for the considerable discrepancy with the $w$-line measurement in Ref.~\cite{Chantler:2012:153001} have been found. 

\section{Acknowledgments}

We thank Vladimir Yerokhin for valuable discussions.
This work was supported by the Russian Science Foundation (Grant No. 17-12-01097).




\end{document}

%% file: tables_v4.tex

\begin{table*}[t]
\centering

\renewcommand{\arraystretch}{1.06}

\caption{\label{tab:Ar_contrib}
Individual contributions to the binding energies of He-like argon evaluated for the LDF, CH, and Coulomb potentials (in eV). For the quasidegenerate states, the contributions to the $H$-matrix elements are listed. See the text for details.
}

\begin{tabular}{l@{\quad}
                l@{\quad}
                S[table-format=-5.5]
                S[table-format=-4.5]
                S[table-format=-2.5]
                S[table-format=-2.5]
                S[table-format=-2.5]
                S[table-format=-2.5]
                S[table-format=-2.5]
                S[table-format=-5.5, table-align-text-post=false]
                }

\hline
 \\[-3mm]

 State  &  Potential  &  {$\Delta E_{\rm Dirac}$}  &  {$\Delta E^{(1)}_{\rm int}$}  &  {$\Delta E^{(2)}_{\rm int}$}  &
                    {$\Delta E^{(\geqslant 3)}_{\rm int}$}  &  {$\Delta E_{\rm 1el}^{\rm QED}$}  &  
                    {$\Delta E_{\rm 2el}^{\rm ScrQED}$}  &  {$\Delta E_{\rm rec}$}  &  {Total} \\[1mm]

\hline
 \\[-3.5mm]

 $(1s1s)_0$       & LDF     &  -8333.8755  &  -213.6274  &  -1.6935  &   0.0282  &  2.1897  &  -0.0311  &  0.1182  &  -8546.8913  \\ 
                  & CH      &  -8247.8460  &  -299.9450  &  -1.3991  &   0.0218  &  2.1577  &   0.0017  &  0.1182  &  -8546.8906  \\ 
                  & Coulomb &  -8854.8313  &   310.2174  &  -4.5781  &   0.0241  &  2.2617  &  -0.1043  &  0.1182  &  -8546.8924  \\
                  &         &              &             &           &           &          &           &          &  -8546.8915$^\ddagger$  \\                  
                  
\hline
 \\[-3.5mm]

 $(1s2s)_0$       & LDF     &  -5177.5501  &  -244.7647  &  -1.4258  &   0.0364  &  1.2349  &   0.0263  &  0.0746  &  -5422.3684  \\
                  & CH      &  -5130.4476  &  -291.1887  &  -2.1248  &   0.0569  &  1.2178  &   0.0435  &  0.0746  &  -5422.3684  \\  
                  & Coulomb &  -5535.4721  &   114.9964  &  -3.2476  &   0.0191  &  1.2834  &  -0.0233  &  0.0746  &  -5422.3694  \\ 
                  &         &              &             &           &           &          &           &          &  -5422.3687$^\ddagger$  \\
                  
\hline
 \\[-3.5mm]

 $(1s2s)_1$       & LDF     &  -5177.5501  &  -265.8997  &  -0.6461  &   0.0108  &  1.2349  &   0.0345  &  0.0748  &  -5442.7409  \\ 
                  & CH      &  -5130.4476  &  -312.3449  &  -1.3187  &   0.0261  &  1.2178  &   0.0516  &  0.0748  &  -5442.7409  \\ 
                  & Coulomb &  -5535.4721  &    92.7048  &  -1.3106  &  -0.0074  &  1.2834  &  -0.0144  &  0.0748  &  -5442.7414  \\ 
                  &         &              &             &           &           &          &           &          &  -5442.7411$^\ddagger$  \\ 
                  
\hline
 \\[-3.5mm]

 $(1s2p_{1/2})_0$ & LDF     &  -5161.5906  &  -262.9518  &  -0.6571  &   0.0110  &  1.0907  &   0.0322  &  0.0661  &  -5423.9995  \\ 
                  & CH      &  -5114.9856  &  -308.8632  &  -1.3673  &   0.0276  &  1.0748  &   0.0481  &  0.0661  &  -5423.9995  \\ 
                  & Coulomb &  -5535.4732  &   112.4409  &  -2.1328  &  -0.0233  &  1.1265  &  -0.0041  &  0.0660  &  -5423.9999  \\
                  &         &              &             &           &           &          &           &          &  -5423.9996$^\ddagger$  \\
                  
\hline
 \\[-3.5mm]

 $(1s2p_{1/2})_1$ & LDF     &  -5161.5906  &  -258.5452  &  -0.9082  &   0.0213  &  1.0907  &   0.0331  &  0.0716  &  -5419.8274  \\ 
                  & CH      &  -5114.9856  &  -304.4669  &  -1.6086  &   0.0384  &  1.0748  &   0.0490  &  0.0716  &  -5419.8274  \\ 
                  & Coulomb &  -5535.4732  &   117.2701  &  -2.8174  &  -0.0022  &  1.1265  &  -0.0033  &  0.0716  &  -5419.8279  \\ 
                  &         &              &             &           &           &          &           &          &  -5419.8276$^\ddagger$  \\
                  
\hline
 \\[-3.5mm]

 $(1s2p_{3/2})_1$ & LDF     &  -5157.4858  &  -253.4334  &  -1.1585  &   0.0315  &  1.0997  &   0.0349  &  0.0770  &  -5410.8346  \\ 
                  & CH      &  -5110.9757  &  -299.2801  &  -1.8393  &   0.0489  &  1.0836  &   0.0510  &  0.0770  &  -5410.8346  \\ 
                  & Coulomb &  -5530.6678  &   122.1620  &  -3.5615  &   0.0211  &  1.1371  &  -0.0030  &  0.0770  &  -5410.8351  \\ 
                  &         &              &             &           &           &          &           &          &  -5410.8350$^\ddagger$  \\

\hline
 \\[-3.5mm]

 $(1s2p_{3/2})_2$ & LDF     &  -5157.4858  &  -263.6602  &  -0.6620  &   0.0113  &  1.0997  &   0.0314  &  0.0660  &  -5420.5997  \\ 
                  & CH      &  -5110.9757  &  -309.4647  &  -1.3843  &   0.0280  &  1.0836  &   0.0475  &  0.0660  &  -5420.5997  \\ 
                  & Coulomb &  -5530.6678  &   110.9057  &  -2.0088  &  -0.0259  &  1.1371  &  -0.0066  &  0.0660  &  -5420.6003  \\ 
                  &         &              &             &           &           &          &           &          &  -5420.5998$^\ddagger$  \\ 
                  
\hline
 \\[-3.5mm]

 off-diag.        & LDF     &      0       &     6.9732  &  -0.3520  &   0.0143  &  0       &   0.0023  &  0.0078  &      6.6456  \\ 
                  & CH      &      0       &     6.9471  &  -0.3264  &   0.0149  &  0       &   0.0023  &  0.0078  &      6.6456  \\ 
                  & Coulomb &      0       &     7.6660  &  -1.0627  &   0.0322  &  0       &   0.0024  &  0.0078  &      6.6457  \\ 
                  &         &              &             &           &           &          &           &          &      6.6455$^\ddagger$  \\ 
                         
\hline

\end{tabular}%

\vspace*{0.7mm}

\hspace*{5mm}
{\small
$^\ddagger$ The higher-order QED correction $\Delta E_{\rm ho}^{\rm QED}$ evaluated according to Ref.~\cite{Drake:1988:586} is added to the total Coulomb value.
}

\end{table*}


\begin{table}[t]
\centering

\renewcommand{\arraystretch}{1.06}

\caption{\label{tab:ho}
Differences between the ground-state binding energies calculated with the effective potentials and the Coulomb potential. LDF and CH stand for the corresponding effective potentials. The higher-order QED correction $\Delta E_{\rm ho}^{\rm QED}$ evaluated according to Ref.~\cite{Drake:1988:586} and the total combined QED-correlation effect obtained in Ref.~\cite{Holmberg:2015:012509} are shown as well. All numbers are given in eV.}

\begin{tabular}{l
                S[table-format=-3.4]
                S[table-format=-3.4]
                S[table-format=-3.4]
                S[table-format=-2.4(1)]                
               }                                

\hline
 \\[-3.5mm]
 
 $Z$  &  {\quad~~ LDF}  &  {\quad\, CH}  &  {~~~~$\Delta E_{\rm ho}^{\rm QED}$}  &  {~~~~QED-corr.}  \\
 
\hline
 \\[-3.5mm]
 
18  &    0.0011   &   0.0018   &    0.0009   &   0.0014(1)   \\

22  &    0.0014   &   0.0023   &    0.0007   &               \\

24  &             &            &    0.0006   &   0.0020(2)   \\

26  &    0.0017   &   0.0028   &    0.0004   &               \\

29  &    0.0020   &   0.0032   &   -0.0001   &               \\

30  &             &            &   -0.0002   &   0.0026(3)   \\

36  &    0.0026   &   0.0041   &   -0.0017   &               \\

\hline

\end{tabular}%

\end{table}


\begin{table*}[t]
\centering

\renewcommand{\arraystretch}{1.06}

\caption{\label{tab:binding}
Binding energies (with the opposite sign) for the $n=1$ and $n=2$ states in He-like ions (in eV).}

\begin{tabular}{l
                S[table-format=6.4(2)]
                S[table-format=6.4(2)]
                S[table-format=6.4(2)]
                S[table-format=6.4(2)]
                S[table-format=6.4(2)]
                S[table-format=6.4(2)]
                S[table-format=6.4(2)]
                }                

\hline
 \\[-3.5mm]

  $Z$    &    {$1s^2\,^1S_0$}     &    {$1s2s\,^1S_0$}     &    {$1s2s\,^3S_1$}     &
              {$1s2p\,^3P_0$}     &    {$1s2p\,^3P_1$}     &    {$1s2p\,^1P_1$}     &    {$1s2p\,^3P_2$}    \\

\hline
    
%
%
%
%

 18 &    8546.8913(9) &   5422.3684(3) &   5442.7409(2) &   5423.9995(2) &
         5423.3548(2) &   5407.3072(5) &   5420.5997(2)  \\ 

 22 &  12874.8344(16) &   8147.1092(5) &   8172.8574(4) &   8149.1938(4) &   
         8147.8946(4) &   8125.1879(7) &   8141.0312(4)  \\ 

 26 &  18105.8782(25) &  11437.8510(9) &  11469.2628(8) &  11440.3139(8) &  
        11438.2967(8) &  11405.4408(9) &  11423.5416(8)  \\ 

 29 &  22630.0566(35) & 14282.8215(12) & 14318.7069(11) & 14285.5112(11) & 
       14283.0606(11) & 14239.0188(12) & 14258.7358(11)  \\ 

 36 &  35232.6631(68) & 22206.1460(28) & 22253.3931(27) & 22209.1433(25) & 
       22206.5427(25) & 22118.1887(25) & 22141.7939(25)  \\     

\hline

\end{tabular}%
\end{table*}


\begin{table*}[t]
\centering

\renewcommand{\arraystretch}{1.06}

\caption{\label{tab:wxyz}
Energies of the x-ray transitions $w(1s2p\,^1P_1 \rightarrow 1s^2\,^1S_0)$, $x(1s2p\,^3P_2 \rightarrow 1s^2\,^1S_0)$, $y(1s2p\,^3P_1 \rightarrow 1s^2\,^1S_0)$, and $z(1s2s\,^3S_1 \rightarrow 1s^2\,^1S_0)$ in heliumlike ions (in eV).
The theoretical (Th.) results are compared with the experimental (Exp.) values. 
The results by Artemyev \textit{et al.}~\cite{Artemyev:2005:062104} are reevaluated using the CODATA 2014 recommended values of the fundamental constants~\cite{Mohr:2016:035009}. }

\begin{tabular}{c
                S[table-format=6.4(2)]
                S[table-format=6.4(2)]
                S[table-format=6.4(2)]
                S[table-format=6.4(2)]
                @{\qquad}l
                @{\quad}r
                @{\quad}l
                }

\hline

  $Z$    &   {$w$}    &
             {$x$}    &
             {$y$}    &
             {$z$}    &
             \hspace*{-4mm} Th./Exp.                         &             
             Year                             &
             \multicolumn{1}{c}{Reference}    \\

\hline
       
   18  &   3139.5842(10)  &   3126.2917(9)   &   3123.5365(9)   &   3104.1504(9)   &  
  Th.  &   2018           &   This work                   \\
       &   3139.5823(4)   &   3126.2898(4)   &   3123.5346(4)   &   3104.1485(4)   &
  Th.  &   2005           &   Artemyev \textit{et al.} \cite{Artemyev:2005:062104} \\     
       &   3139.559(16)   &                  &                  &                  &   
  Th.  &   2018           &   Machado \textit{et al.} \cite{Machado:2018:032517}  \\       
       &   3139.5927(76)  &                  &                  &                  &   
  Exp. &   2018           &   Machado \textit{et al.} \cite{Machado:2018:032517}  \\
       &   3139.581(9)    &                  &                  &                  &   
  Exp. &   2014           &   Kubi{\v c}ek \textit{et al.} \cite{Kubicek:2014:032508}   \\
       &                  &                  &                  &   3104.1605(77)  &   
  Exp. &   2012           &   Amaro \textit{et al.} \cite{Amaro:2012:043005}     \\
       &   3139.583(6)    &                  &                  &                  &   
  Exp. &   2007           &   Bruhns \textit{et al.} \cite{Bruhns:2007:113001}    \\
       &   3139.552(36)   &   3126.283(36)   &   3123.521(36)   &                  &   
  Exp. &   1984           &   Deslattes \textit{et al.} \cite{Deslattes:1984:L689}  \\
       &   3139.6(3)      &   3126.4(4)      &   3123.6(3)      &                  &   
  Exp. &   1983           &   Briand \textit{et al.} \cite{Briand:1983:1413}     \\

\hline

   22  &   4749.6465(16)  &   4733.8032(16)  &   4726.9399(16)  &   4701.9771(16)  &   
  Th.  &   2018           &   This work                   \\
       &   4749.6444(6)   &   4733.8011(6)   &   4726.9376(6)   &   4701.9749(7)   &   
  Th.  &   2005           &   Artemyev \textit{et al.} \cite{Artemyev:2005:062104}  \\ 
       &   4749.6521      &   4733.8167      &   4726.9524      &   4701.9846      & 
  Th.  &   2000           &   Cheng {\footnotesize\&} Chen \cite{Cheng:2000:044503} \\          
       &                  &   4733.83(13)    &   4727.07(10)    &   4702.078(72)   &   
  Exp. &   2014           &   Payne \textit{et al.} \cite{Payne:2014:185001}     \\
       &   4749.852(72)   &                  &                  &                  &   
  Exp. &   2012           &   Chantler \textit{et al.} \cite{Chantler:2012:153001}   \\
       &   4749.73(17)    &                  &                  &                  &   
  Exp. &   1989           &   Beiersdorfer \textit{et al.} \cite{Beiersdorfer:1989:150}  \\

\hline

   26  &   6700.4373(25)  &   6682.3366(25)  &   6667.5814(25)  &   6636.6154(25)  &   
  Th.  &   2018           &   This work                    \\
       &   6700.4351(11)  &   6682.3343(11)  &   6667.5790(12)  &   6636.6130(13)  &   
  Th.  &   2005           &   Artemyev \textit{et al.} \cite{Artemyev:2005:062104}  \\  
       &   6700.4505      &   6682.3613      &   6667.6043      &   6636.6314      & 
  Th.  &   2000           &   Cheng {\footnotesize\&} Chen \cite{Cheng:2000:044503} \\       
       &   6700.441(49)   &                  &                  &                  &   
  Exp. &   2014           &   Kubi{\v c}ek \textit{et al.} \cite{Kubicek:2014:032508}   \\
       &   6700.549(70)   &                  &   6667.671(69)   &                  &   
  Exp. &   2013           &   Rudolph \textit{et al.} \cite{Rudolph:2013:103002}   \\
       &   6700.73(20)    &                  &                  &                  &   
  Exp. &   1989           &   Beiersdorfer \textit{et al.} \cite{Beiersdorfer:1989:150} \\
       &   6700.76(36)    &                  &                  &                  &   
  Exp. &   1988           &   Aglitsky \textit{et al.} \cite{Aglitsky:1988:136}     \\
       &   6700.9(3)      &   6682.7(3)      &   6667.5(3)      &                  &
  Exp. &   1984           &   Briand \textit{et al.} \cite{Briand:1984:3143}        \\

\hline

   29  &   8391.0378(34)  &   8371.3208(34)  &   8346.9960(34)  &   8311.3497(34)  &   
  Th.  &   2018           &   This work                    \\
       &   8391.0354(17)  &   8371.3186(17)  &   8346.9934(17)  &   8311.3472(20)  &   
  Th.  &   2005           &   Artemyev \textit{et al.} \cite{Artemyev:2005:062104}   \\   
       &   8391.0590      &   8371.3560      &   8347.0283      &   8311.3739      & 
  Th.  &   2000           &   Cheng {\footnotesize\&} Chen \cite{Cheng:2000:044503} \\       
       &   8390.82(15)    &   8371.17(15)    &   8346.99(15)    &   8310.83(15)    &   
  Exp. &   2015           &   Beiersdorfer {\footnotesize\&} Brown \cite{Beiersdorfer:2015:032514}  \\
       &   8391.03(40)    &                  &                  &                  &   
  Exp. &   1988           &   Aglitsky \textit{et al.} \cite{Aglitsky:1988:136}      \\

\hline

   36  &   13114.4744(62) &   13090.8692(62) &   13026.1204(62) &   12979.2700(62) &   
  Th.  &   2018           &   This work                    \\
       &   13114.4712(41) &   13090.8664(41) &   13026.1172(42) &   12979.2663(56) &   
  Th.  &   2005           &   Artemyev \textit{et al.} \cite{Artemyev:2005:062104}  \\ 
       &   13114.5051     &   13090.9214     &   13026.1660     &   12979.3055     &   
  Th.  &   2000           &   Cheng {\footnotesize\&} Chen \cite{Cheng:2000:044503} \\        
       &   13114.47(14)   &                  &   13026.15(14)   &                  &   
  Exp. &   2015           &   Epp \textit{et al.} \cite{Epp:2015:020502_R}     \\
       &   13114.68(36)   &   13091.17(37)   &   13026.29(36)   &   12979.63(41)   &   
  Exp. &   1996           &   Widmann \textit{et al.} \cite{Widmann:1996:2200}     \\
       &   13115.45(30)   &                  &   13026.80(30)   &                  &
  Exp. &   1986           &   Indelicato \textit{et al.} \cite{Indelicato:1986:249} \\

\hline

\end{tabular}%
\end{table*}


%% file: he_transition_v9.bbl
\begin{thebibliography}{10}

\bibitem{Schweppe:1991:1434}
J.{~}Schweppe, A.{~}Belkacem, L.{~}Blumenfeld, N.{~}Claytor, B.{~}Feinberg,
  H.{~}Gould, V.~E.{~}Kostroun, L.{~}Levy, S.{~}Misawa, J.~R.{~}Mowat, and
  M.~H.{~}Prior,
\newblock Phys. Rev. Lett. {\bf 66},~1434 (1991).

\bibitem{Stoehlker:2000:3109}
{\relax Th}.{~}St\"ohlker, P.~H.{~}Mokler, F.{~}Bosch, R.~W.{~}Dunford,
  F.{~}Franzke, O.{~}Klepper, C.{~}Kozhuharov, T.{~}Ludziejewski, F.{~}Nolden,
  H.{~}Reich, P.{~}Rymuza, Z.{~}Stachura, M.{~}Steck, P.{~}Swiat, and
  A.{~}Warczak,
\newblock Phys. Rev. Lett. {\bf 85},~3109 (2000).

\bibitem{Brandau:2003:073202}
C.{~}Brandau, C.{~}Kozhuharov, A.{~}M\"uller, W.{~}Shi, S.{~}Schippers,
  T.{~}Bartsch, S.{~}B\"ohm, C.{~}B\"ohme, A.{~}Hoffknecht, H.{~}Knopp,
  N.{~}Gr\"un, W.{~}Scheid, T.{~}Steih, F.{~}Bosch, B.{~}Franzke,
  P.~H.{~}Mokler, F.{~}Nolden, M.{~}Steck, T.{~}St\"ohlker, and Z.{~}Stachura,
\newblock Phys. Rev. Lett. {\bf 91},~073202 (2003).

\bibitem{Draganic:2003:183001}
I.{~}Dragani\'c, J.~R.{~}{Crespo L\'opez-Urrutia}, R.{~}DuBois, S.{~}Fritzsche,
  V.~M.{~}Shabaev, R.{~}Soria~Orts, I.~I.{~}Tupitsyn, Y.{~}Zou, and
  J.{~}Ullrich,
\newblock Phys. Rev. Lett. {\bf 91},~183001 (2003).

\bibitem{Gumberidze:2004:203004}
A.{~}Gumberidze, {\relax Th}.{~}St\"ohlker, D.{~}Bana\'s, K.{~}Beckert,
  P.{~}Beller, H.~F.{~}Beyer, F.{~}Bosch, X.{~}Cai, S.{~}Hagmann,
  C.{~}Kozhuharov, D.{~}Liesen, F.{~}Nolden, X.{~}Ma, P.~H.{~}Mokler,
  A.{~}{Or{\v s}i\'c-Muthig}, M.{~}Steck, D.{~}Sierpowski, S.{~}Tashenov,
  A.{~}Warczak, and Y.{~}Zou,
\newblock Phys. Rev. Lett. {\bf 92},~203004 (2004).

\bibitem{Gumberidze:2005:223001}
A.{~}Gumberidze, {\relax Th}.{~}St\"ohlker, D.{~}Bana\'s, K.{~}Beckert,
  P.{~}Beller, H.~F.{~}Beyer, F.{~}Bosch, S.{~}Hagmann, C.{~}Kozhuharov,
  D.{~}Liesen, F.{~}Nolden, X.{~}Ma, P.~H.{~}Mokler, M.{~}Steck,
  D.{~}Sierpowski, and S.{~}Tashenov,
\newblock Phys. Rev. Lett. {\bf 94},~223001 (2005).

\bibitem{Beiersdorfer:2005:233003}
P.{~}Beiersdorfer, H.{~}Chen, D.~B.{~}Thorn, and E.{~}Tr\"abert,
\newblock Phys. Rev. Lett. {\bf 95},~233003 (2005).

\bibitem{Mackel:2011:143002}
V.{~}M\"ackel, R.{~}Klawitter, G.{~}Brenner, J.~R.{~}{Crespo L\'opez-Urrutia},
  and J.{~}Ullrich,
\newblock Phys. Rev. Lett. {\bf 107},~143002 (2011).

\bibitem{Persson:1996:204}
H.{~}Persson, S.{~}Salomonson, P.{~}Sunnergren, and I.{~}Lindgren,
\newblock Phys. Rev. Lett. {\bf 76},~204 (1996).

\bibitem{Yerokhin:1997:361}
V.~A.{~}Yerokhin, A.~N.{~}Artemyev, and V.~M.{~}Shabaev,
\newblock Phys. Lett. A {\bf 234},~361 (1997).

\bibitem{Yerokhin:2006:253004}
V.~A.{~}Yerokhin, P.{~}Indelicato, and V.~M.{~}Shabaev,
\newblock Phys. Rev. Lett. {\bf 97},~253004 (2006).

\bibitem{Artemyev:2007:173004}
A.~N.{~}Artemyev, V.~M.{~}Shabaev, I.~I.{~}Tupitsyn, G.{~}Plunien, and
  V.~A.{~}Yerokhin,
\newblock Phys. Rev. Lett. {\bf 98},~173004 (2007).

\bibitem{Kozhedub:2008:032501}
Y.~S.{~}Kozhedub, O.~V.{~}Andreev, V.~M.{~}Shabaev, I.~I.{~}Tupitsyn,
  C.{~}Brandau, C.{~}Kozhuharov, G.{~}Plunien, and T.{~}St\"ohlker,
\newblock Phys. Rev. A {\bf 77},~032501 (2008).

\bibitem{Kozhedub:2010:042513}
Y.~S.{~}Kozhedub, A.~V.{~}Volotka, A.~N.{~}Artemyev, D.~A.{~}Glazov,
  G.{~}Plunien, V.~M.{~}Shabaev, I.~I.{~}Tupitsyn, and {\relax
  Th}.{~}St\"ohlker,
\newblock Phys. Rev. A {\bf 81},~042513 (2010).

\bibitem{Sapirstein:2011:012504}
J.{~}Sapirstein and K.~T.{~}Cheng,
\newblock Phys. Rev. A {\bf 83},~012504 (2011).

\bibitem{Artemyev:2013:032518}
A.~N.{~}Artemyev, V.~M.{~}Shabaev, I.~I.{~}Tupitsyn, G.{~}Plunien,
  A.{~}Surzhykov, and S.{~}Fritzsche,
\newblock Phys. Rev. A {\bf 88},~032518 (2013).

\bibitem{Yerokhin:2015:033103}
V.~A.{~}Yerokhin and V.~M.{~}Shabaev,
\newblock J. Phys. Chem. Ref. Data {\bf 44},~033103 (2015).

\bibitem{Yerokhin:2018:052509}
V.~A.{~}Yerokhin,
\newblock Phys. Rev. A {\bf 97},~052509 (2018).

\bibitem{Drake:1988:586}
G.~W.{~}Drake,
\newblock Can. J. Phys. {\bf 66},~586 (1988).

\bibitem{Johnson:1992:R2197}
W.~R.{~}Johnson and J.{~}Sapirstein,
\newblock Phys. Rev. A {\bf 46},~R2197 (1992).

\bibitem{Chen:1993:3692}
M.~H.{~}Chen, K.~T.{~}Cheng, and W.~R.{~}Johnson,
\newblock Phys. Rev. A {\bf 47},~3692 (1993).

\bibitem{Plante:1994:3519}
D.~R.{~}Plante, W.~R.{~}Johnson, and J.{~}Sapirstein,
\newblock Phys. Rev. A {\bf 49},~3519 (1994).

\bibitem{Cheng:1994:247}
K.~T.{~}Cheng, M.~H.{~}Chen, W.~R.{~}Johnson, and J.{~}Sapirstein,
\newblock Phys. Rev. A {\bf 50},~247 (1994).

\bibitem{Indelicato:1995:1132}
P.{~}Indelicato,
\newblock Phys. Rev. A {\bf 51},~1132 (1995).

\bibitem{Cheng:2000:044503}
K.~T.{~}Cheng and M.~H.{~}Chen,
\newblock Phys. Rev. A {\bf 61},~044503 (2000).

\bibitem{Artemyev:2005:062104}
A.~N.{~}Artemyev, V.~M.{~}Shabaev, V.~A.{~}Yerokhin, G.{~}Plunien, and
  G.{~}Soff,
\newblock Phys. Rev. A {\bf 71},~062104 (2005).

\bibitem{Deslattes:1984:L689}
R.~D.{~}Deslattes, H.~F.{~}Beyer, and F.{~}Folkmann,
\newblock J. Phys. B: At. Mol. Phys. {\bf 17},~L689 (1984).

\bibitem{Beiersdorfer:1989:150}
P.{~}Beiersdorfer, M.{~}Bitter, S.{~}{von Goeler}, and K.~W.{~}Hill,
\newblock Phys. Rev. A {\bf 40},~150 (1989).

\bibitem{Chantler:2000:042501}
C.~T.{~}Chantler, D.{~}Paterson, L.~T.{~}Hudson, F.~G.{~}Serpa,
  J.~D.{~}Gillaspy, and E.{~}Tak\'acs,
\newblock Phys. Rev. A {\bf 62},~042501 (2000).

\bibitem{Howie:1994:4390}
D.~J.~H.{~}Howie, W.~A.{~}Hallett, E.~G.{~}Myers, D.~D.{~}Dietrich, and
  J.~D.{~}Silver,
\newblock Phys. Rev. A {\bf 49},~4390 (1994).

\bibitem{Kukla:1995:1905}
K.~W.{~}Kukla, A.~E.{~}Livingston, J.{~}Suleiman, H.~G.{~}Berry,
  R.~W.{~}Dunford, D.~S.{~}Gemmell, E.~P.{~}Kanter, S.{~}Cheng, and
  L.~J.{~}Curtis,
\newblock Phys. Rev. A {\bf 51},~1905 (1995).

\bibitem{Redshaw:2001:23002}
M.{~}Redshaw and E.~G.{~}Myers,
\newblock Phys. Rev. Lett. {\bf 88},~023002 (2001).

\bibitem{Chantler:2012:153001}
C.~T.{~}Chantler, M.~N.{~}Kinnane, J.~D.{~}Gillaspy, L.~T.{~}Hudson,
  A.~T.{~}Payne, L.~F.{~}Smale, A.{~}Henins, J.~M.{~}Pomeroy, J.~N.{~}Tan,
  J.~A.{~}Kimpton, E.{~}Takacs, and K.{~}Makonyi,
\newblock Phys. Rev. Lett. {\bf 109},~153001 (2012).

\bibitem{Chantler:2014:123037}
C.~T.{~}Chantler, A.~T.{~}Payne, J.~D.{~}Gillaspy, L.~T.{~}Hudson,
  L.~F.{~}Smale, A.{~}Henins, J.~A.{~}Kimpton, and E.{~}Takacs,
\newblock New J. Phys. {\bf 16},~123037 (2014).

\bibitem{Rudolph:2013:103002}
J.~K.{~}Rudolph, S.{~}Bernitt, S.~W.{~}Epp, R.{~}Steinbr\"ugge, C.{~}Beilmann,
  G.~V.{~}Brown, S.{~}Eberle, A.{~}Graf, Z.{~}Harman, N.{~}Hell,
  M.{~}Leutenegger, A.{~}M\"uller, K.{~}Schlage, H.-C.{~}Wille, H.{~}Yava{\c
  s}, J.{~}Ullrich, and J.~R.{~}{Crespo L\'opez-Urrutia},
\newblock Phys. Rev. Lett. {\bf 111},~103002 (2013).

\bibitem{Schlesser:2013:022503}
S.{~}Schlesser, S.{~}Boucard, D.~S.{~}Covita, J.~M.~F.{~}{dos Santos},
  H.{~}Fuhrmann, D.{~}Gotta, A.{~}Gruber, M.{~}Hennebach, A.{~}Hirtl,
  P.{~}Indelicato, E.-O.{~}Le~Bigot, L.~M.{~}Simons, L.{~}Stingelin,
  M.{~}Trassinelli, J.~F. C.~A.{~}Veloso, A.{~}Wasser, and J.{~}Zmeskal,
\newblock Phys. Rev. A {\bf 88},~022503 (2013).

\bibitem{Kubicek:2014:032508}
K.{~}Kubi{\v c}ek, P.~H.{~}Mokler, V.{~}M\"ackel, J.{~}Ullrich, and
  J.~R.{~}{Crespo L\'opez-Urrutia},
\newblock Phys. Rev. A {\bf 90},~032508 (2014).

\bibitem{Epp:2015:020502_R}
S.~W.{~}Epp, R.{~}Steinbr\"ugge, S.{~}Bernitt, J.~K.{~}Rudolph, C.{~}Beilmann,
  H.{~}Bekker, A.{~}M\"uller, O.~O.{~}Versolato, H.-C.{~}Wille, H.{~}Yava{\c
  s}, J.{~}Ullrich, and J.~R.{~}{Crespo L\'opez-Urrutia},
\newblock Phys. Rev. A {\bf 92},~020502(R) (2015).

\bibitem{Beiersdorfer:2015:032514}
P.{~}Beiersdorfer and G.~V.{~}Brown,
\newblock Phys. Rev. A {\bf 91},~032514 (2015).

\bibitem{Machado:2018:032517}
J.{~}Machado, C.~I.{~}Szabo, J.~P.{~}Santos, P.{~}Amaro, M.{~}Guerra,
  A.{~}Gumberidze, G.{~}Bian, J.~M.{~}Isac, and P.{~}Indelicato,
\newblock Phys. Rev. A {\bf 97},~032517 (2018).

\bibitem{Epp:2013:159301}
S.~W.{~}Epp,
\newblock Phys. Rev. Lett. {\bf 110},~159301 (2013).

\bibitem{Malyshev:2018:085001}
A.~V.{~}Malyshev, R.~V.{~}Popov, V.~M.{~}Shabaev, and N.~A.{~}Zubova,
\newblock J. Phys. B: At. Mol. Opt. Phys. {\bf 51},~085001 (2018).

\bibitem{Gabriel:1972:99}
A.~H.{~}Gabriel,
\newblock Mon. Not. R. Astron. Soc. {\bf 160},~99 (1972).

\bibitem{TTGF}
V.~M.{~}Shabaev,
\newblock Phys. Rep. {\bf 356},~119 (2002).

\bibitem{Sapirstein:2001:022502}
J.{~}Sapirstein and K.~T.{~}Cheng,
\newblock Phys. Rev. A {\bf 64},~022502 (2001).

\bibitem{Yerokhin:2007:062501}
V.~A.{~}Yerokhin, A.~N.{~}Artemyev, and V.~M.{~}Shabaev,
\newblock Phys. Rev. A {\bf 75},~062501 (2007).

\bibitem{Malyshev:2017:022512}
A.~V.{~}Malyshev, D.~A.{~}Glazov, A.~V.{~}Volotka, I.~I.{~}Tupitsyn,
  V.~M.{~}Shabaev, G.{~}Plunien, and {\relax Th}.{~}St\"ohlker,
\newblock Phys. Rev. A {\bf 96},~022512 (2017).

\bibitem{Sapirstein:2001:032506}
J.{~}Sapirstein and K.~T.{~}Cheng,
\newblock Phys. Rev. A {\bf 63},~032506 (2001).

\bibitem{Glazov:2006:330}
D.~A.{~}Glazov, A.~V.{~}Volotka, V.~M.{~}Shabaev, I.~I.{~}Tupitsyn, and
  G.{~}Plunien,
\newblock Phys. Lett. A {\bf 357},~330 (2006).

\bibitem{Volotka:2014:253004}
A.~V.{~}Volotka, D.~A.{~}Glazov, V.~M.{~}Shabaev, I.~I.{~}Tupitsyn, and
  G.{~}Plunien,
\newblock Phys. Rev. Lett. {\bf 112},~253004 (2014).

\bibitem{Sapirstein:2002:042501}
J.{~}Sapirstein and K.~T.{~}Cheng,
\newblock Phys. Rev. A {\bf 66},~042501 (2002).

\bibitem{Chen:2006:042510}
M.~H.{~}Chen, K.~T.{~}Cheng, W.~R.{~}Johnson, and J.{~}Sapirstein,
\newblock Phys. Rev. A {\bf 74},~042510 (2006).

\bibitem{Sapirstein:2003:022512}
J.{~}Sapirstein and K.~T.{~}Cheng,
\newblock Phys. Rev. A {\bf 67},~022512 (2003).

\bibitem{Sapirstein:2006:042513}
J.{~}Sapirstein and K.~T.{~}Cheng,
\newblock Phys. Rev. A {\bf 74},~042513 (2006).

\bibitem{Artemyev:2000:022116}
A.~N.{~}Artemyev, T.{~}Beier, G.{~}Plunien, V.~M.{~}Shabaev, G.{~}Soff, and
  V.~A.{~}Yerokhin,
\newblock Phys. Rev. A {\bf 62},~022116 (2000).

\bibitem{Lindgren:2001:062505}
I.{~}Lindgren, B.{~}\AA{}s\'en, S.{~}Salomonson, and
  A.-M.{~}{M\aa{}rtensson-Pendrill},
\newblock Phys. Rev. A {\bf 64},~062505 (2001).

\bibitem{Andreev:2004:062505}
O.~Y.{~}Andreev, L.~N.{~}Labzowsky, G.{~}Plunien, and G.{~}Soff,
\newblock Phys. Rev. A {\bf 69},~062505 (2004).

\bibitem{Shabaev:2005:062105}
V.~M.{~}Shabaev, I.~I.{~}Tupitsyn, K.{~}Pachucki, G.{~}Plunien, and
  V.~A.{~}Yerokhin,
\newblock Phys. Rev. A {\bf 72},~062105 (2005).

\bibitem{Mohr:2016:035009}
P.~J.{~}Mohr, D.~B.{~}Newell, and B.~N.{~}Taylor,
\newblock Rev. Mod. Phys. {\bf 88},~035009 (2016).

\bibitem{Bruhns:2007:113001}
H.{~}Bruhns, J.{~}Braun, K.{~}Kubi{\v c}ek, J.~R.{~}{Crespo L\'opez-Urrutia},
  and J.{~}Ullrich,
\newblock Phys. Rev. Lett. {\bf 99},~113001 (2007).

\bibitem{Aglitsky:1988:136}
E.~V.{~}Aglitsky, P.~S.{~}Antsiferov, S.~L.{~}Mandelstam, A.~M.{~}Panin,
  U.~I.{~}Safronova, S.~A.{~}Ulitin, and L.~A.{~}Vainshtein,
\newblock Phys. Scr. {\bf 38},~136 (1988).

\bibitem{Widmann:1996:2200}
K.{~}Widmann, P.{~}Beiersdorfer, V.{~}Decaux, and M.{~}Bitter,
\newblock Phys. Rev. A {\bf 53},~2200 (1996).

\bibitem{Yerokhin:2003:203}
V.~A.{~}Yerokhin, P.{~}Indelicato, and V.~M.{~}Shabaev,
\newblock Eur. Phys. J. D {\bf 25},~203 (2003).

\bibitem{Yerokhin:2008:062510}
V.~A.{~}Yerokhin, P.{~}Indelicato, and V.~M.{~}Shabaev,
\newblock Phys. Rev. A {\bf 77},~062510 (2008).

\bibitem{Sanders:1969:84}
F.~C.{~}Sanders and C.~W.{~}Scherr,
\newblock Phys. Rev. {\bf 181},~84 (1969).

\bibitem{Aashamar:1970:3324}
K.{~}Aashamar, G.{~}Lyslo, and J.{~}Midtdal,
\newblock J. Chem. Phys. {\bf 52},~3324 (1970).

\bibitem{Tupitsyn:2003:022511}
I.~I.{~}Tupitsyn, V.~M.{~}Shabaev, J.~R.{~}{Crespo L\'opez-Urrutia},
  I.{~}Dragani\'c, R.{~}Soria~Orts, and J.{~}Ullrich,
\newblock Phys. Rev. A {\bf 68},~022511 (2003).

\bibitem{Glazov:2017:46}
D.~A.{~}Glazov, A.~V.{~}Malyshev, A.~V.{~}Volotka, V.~M.{~}Shabaev,
  I.~I.{~}Tupitsyn, and G.{~}Plunien,
\newblock Nucl. Instrum. Methods Phys. Res., Sect. B {\bf 408},~46 (2017).

\bibitem{Brouder:2012:2256}
C.{~}Brouder, G.~H.{~}Duchamp, F.{~}Patras, and G.~Z.{~}T\'oth,
\newblock Int. J. Quantum Chem. {\bf 112},~2256 (2012).

\bibitem{Holmberg:2015:012509}
J.{~}Holmberg, S.{~}Salomonson, and I.{~}Lindgren,
\newblock Phys. Rev. A {\bf 92},~012509 (2015).

\bibitem{Amaro:2012:043005}
P.{~}Amaro, S.{~}Schlesser, M.{~}Guerra, E.-O.{~}Le~Bigot, J.-M.{~}Isac,
  P.{~}Travers, J.~P.{~}Santos, C.~I.{~}Szabo, A.{~}Gumberidze, and
  P.{~}Indelicato,
\newblock Phys. Rev. Lett. {\bf 109},~043005 (2012).

\bibitem{Briand:1983:1413}
J.~P.{~}Briand, J.~P.{~}Moss\'e, P.{~}Indelicato, P.{~}Chevallier,
  D.{~}{Girard-Vernhet}, A.{~}Chetioui, M.~T.{~}Ramos, and J.~P.{~}Desclaux,
\newblock Phys. Rev. A {\bf 28},~1413 (1983).

\bibitem{Payne:2014:185001}
A.~T.{~}Payne, C.~T.{~}Chantler, M.~N.{~}Kinnane, J.~D.{~}Gillaspy,
  L.~T.{~}Hudson, L.~F.{~}Smale, A.{~}Henins, J.~A.{~}Kimpton, and E.{~}Takacs,
\newblock J. Phys. B: At. Mol. Opt. Phys. {\bf 47},~185001 (2014).

\bibitem{Briand:1984:3143}
J.~P.{~}Briand, M.{~}Tavernier, R.{~}Marrus, and J.~P.{~}Desclaux,
\newblock Phys. Rev. A {\bf 29},~3143 (1984).

\bibitem{Indelicato:1986:249}
P.{~}Indelicato, J.~P.{~}Briand, M.{~}Tavernier, and D.{~}Liesen,
\newblock Z. Phys. D {\bf 2},~249 (1986).

\bibitem{Angeli:2013:69}
I.{~}Angeli and K.~P.{~}Marinova,
\newblock At. Data Nucl. Data Tables {\bf 99},~69 (2013).

\end{thebibliography}
